\documentclass[twocolumn,preprintnumbers,letterpaper,superscriptaddress,showpacs,pra]{revtex4}
\usepackage{amsmath}
\usepackage{graphicx}
\usepackage{dcolumn}
\usepackage{bm}
\usepackage{subfigure}

\begin{document}

\title{Non-equilibrium Enhancement of Cooper Pairing in Cold Fermion Systems}
\author{Andrew Robertson and Victor M. Galitski}
\affiliation{Joint Quantum Institute and Condensed Matter Theory Center, Department of
Physics, University of Maryland, College Park, Maryland 20742-4111, USA}

\begin{abstract}
Non-equilibrium stimulation of superfluidity in trapped Fermi gases is
discussed by analogy to the work of Eliashberg [G.~M. ~Eliashberg, in
``Nonequilibrium Superconductivity,'' edited by D.~N.~Langenberg and
A.~I.~Larkin (North-Holland, New York, 1986)] on the microwave enhancement
of superconductivity. Optical excitation of the fermions balanced by heat
loss due to thermal contact with a boson bath and/or evaporative cooling
enables stationary non-equilibrium states to exist. Such a state manifests
as a shift of the quasiparticle spectrum to higher energies and this
effectively raises the pairing transition temperature. As an illustration,
we calculate the effective enhancement of Cooper pairing and superfluidity
in both the normal and superfluid phases for a mixture of $^{87}$Rb and $%
^{6} $Li in the limit of small departure from equilibrium. It is argued that
in experiment the desirable effect is not limited to such small
perturbations and the effective enhancement of the pairing temperature may
be quite large.%
\end{abstract}

\pacs{03.75.Hh, 67.85.Lm, 67.10.Db}
\date{\today }
\maketitle

\section{Introduction}

The difficulty of observing quantum coherent phases in cold gases highlights
the need to overcome low transition temperatures. In addition to
Bose-Einstein condensation, there has also been recent interest in the
generation of Fermi superfluids through the BCS pairing mechanism~\cite%
{CABCS}. This phenomenon is more difficult to observe due to prohibitively
low transition temperatures \cite{Holland01,Stoof96} though the problem may
be partially surmounted by use of Feshbach resonances \cite%
{CABCS,Zwierlein06,Stoof99}. Non-equilibrium effects can also be used to
control and effectively cool such systems. However, this is an unexplored
area of research by comparison.

Since the 1960's, it was known that superconductivity could be stimulated by
radiation in microbridges \cite{Wyatt66}. In 1970, Eliashberg explained this
effect as an amplification of the gap parameter by means of a stationary
nonequilibrium shift in the quasiparticle spectrum to higher energies
brought on by the radiation \cite{Eliashberg70,Larkin86}. Over the next
decade, his theory found experimental acceptance through the enhancement of
critical currents and temperatures in Josephson junctions \cite{Tredwell75}
and thin films \cite{Pals79}. At the same time, other nonequilibrium
stimulation methods were developed \cite{Chang78} with more recent reports
of enhancements of the superconducting critical temperature by up to several
times its equilibrium value \cite{Blamire91,Heslinga93}. With the present
interest in the application of the BCS model of superconductivity to trapped
atomic Fermi gases \cite{Carr04,Heslinga93,Bruun00}, nonequilibrium effects
represent an attractive way to magnify the quantum properties of these types
of superfluids.

As in superconductors, the BCS order parameter $\Delta _{0}$ for cold
fermionic gases obeys a self-consistency equation \cite%
{Sademelo93,Stoof99,Sheehy06}%
\begin{equation}
\frac{\Delta _{0}}{\lambda }=\Delta _{0}\sum_{\mathbf{k}}\left[ \frac{1-2n_{%
\mathbf{k}}}{2\sqrt{\xi _{\mathbf{k}}^{2}+\Delta _{0}^{2}}}-\frac{1}{2\xi _{%
\mathbf{k}}}\right] ,  \label{GapEq}
\end{equation}%
where $\xi _{\mathbf{k}}=\varepsilon _{\mathbf{k}}-\varepsilon _{\mathrm{F}}$
is the quasiparticle dispersion centered on the Fermi energy $\varepsilon _{%
\mathrm{F}}$, $\Delta _{0}$ is the BCS gap, and $n_{\mathbf{k}}$ is the
quasiparticle distribution function. The constant $\lambda $ has the form $%
\lambda =-4\pi a_{\uparrow \downarrow }/m_{F}$ where $a_{\uparrow \downarrow
}$ is the negative s-wave scattering length for collisions between hyperfine
states and $m_{F}$ is the mass. At equilibrium, $n_{\mathbf{k}}^{\mathrm{(FD)%
}}$ is the Fermi-Dirac distribution function, and the only way to increase $%
\Delta _{0}$ is either to increase the interaction strength or to lower the
temperature. However, there exists a wide class of stationary nonequilibrium
distributions, $n_{\mathbf{k}}$, such that Eq.~(\ref{GapEq}) is still valid
and has solutions with enhanced order parameters. Indeed, if a
quasistationary non-equilibrium distribution is created that is different
from the canonical Fermi-Dirac function, $\delta n_{\mathbf{k}}=n_{\mathbf{k}%
}-n_{\mathbf{k}}^{\mathrm{(FD)}}$, then according to the weak-coupling BCS
Eq.~(\ref{GapEq}), it effectively renormalizes the pairing interaction and
transition temperature as follows:
\begin{equation}
{\frac{1}{\lambda _{\mathrm{eff}}}}={\frac{1}{\lambda }}+\sum_{\mathbf{k}}%
\frac{\delta n_{\mathbf{k}}}{E_{\mathbf{k}}}\equiv {\frac{1}{\lambda }}-\nu
(\varepsilon _{\mathrm{F}})\chi \,\,\mbox{ and }\,\,T_{c}^{\mathrm{(eff)}%
}=T_{c}^{(0)}e^{\chi },  \label{renorm}
\end{equation}%
where here and below $E_{\mathbf{k}}=\sqrt{\xi _{\mathbf{k}}^{2}+\Delta
_{0}^{2}}$, $\nu (\varepsilon _{\mathrm{F}})$ is the density of states at
the Fermi level, $T_{c}^{(0)}\sim \varepsilon _{\mathrm{F}}\exp \left\{ -1/%
\left[ \nu (\varepsilon _{\mathrm{F}})\lambda \right] \right\} $ is the
weak-coupling BCS transition temperature in equilibrium, and we also
introduced the dimensionless parameter $\chi =-\sum_{\mathbf{k}}\delta n_{%
\mathbf{k}}/\left[ \nu (\varepsilon _{\mathrm{F}})E_{\mathbf{k}}\right] $.
For many non-equilibrium distributions, $\chi >0$, and this yields an
effective enhancement of $T_{c}$ and/or $\Delta $. We note that even though
our theory below and that of Eliashberg are limited to small deviations from
equilibrium, with $|\chi| \ll 1$, this does not imply a limitation in
experiment, where this parameter can be large. For such large deviations
from equilibrium, the weak-coupling BCS approach and Eqs.~(\ref{GapEq}) and (%
\ref{renorm}) may not be quantitatively applicable, but the tendency to
enhance pairing may remain. Therefore the proposed underlying mechanism may lead to
substantial enhancement of fermion pairing and superfluidity. We also
emphasize here that cold atom systems offer more control in creating and
manipulating non-equilibrium many-body quantum states than that available in
solids. Specifically, we will show that while it was impossible to drive a
metal from the normal to the superconducting phase by irradiation, it is
indeed possible to drive the equivalent transition in cold gas by utilizing
this additional control.

\begin{figure}[t]
\includegraphics[width=2.5in]{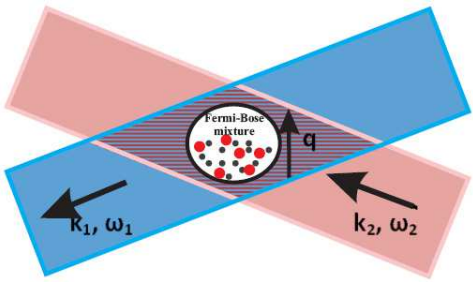}
\caption{{\protect\footnotesize Bragg Potential: \ A moving lattice with
wave vector }$\mathbf{q}=\mathbf{k}_{2}-\mathbf{k}_{1}$%
{\protect\footnotesize \ can be formed in the region of the Bose-Fermi
mixture through the interference of two lasers with differing wave vectors
and frequencies. By adjusting the parameters of this non-equilibrium
perturbation, one can achieve states with enhanced superfluidity.}}
\label{Brgpic}
\end{figure}

In this paper, we propose a theory of nonequilibrium stimulation of fermion
pairing by considering the effect of Bragg pulses \cite{Blakie02,Rey05} as
shown schematically in Fig.~\ref{Brgpic} on a harmonically trapped gas of fermions in the
Thomas-Fermi approximation \cite{Butts97}. The heating induced by the
external perturbation is dumped into an isothermal bath of trapped bosons
via collisions, but this is not necessary in general. The pairing
enhancement is calculated for a typical mixture of $^{87}$Rb and $^{6}$Li.
It depends on the state of the gas at equilibrium. In the superfluid phase,
Eliashberg's requirement that the frequency of the perturbation be less than
twice the equilibrium gap ($\hbar \omega <2\Delta _{0}$) ensures that the
pulse does not effectively heat the system by producing more quasiparticles
with energies $\varepsilon \sim \Delta _{0}$. Though this requirement cannot
be satisfied in the normal phase, where $\Delta _{0}=0$, the independent
tunability of both the momentum and energy of the Bragg pulse allows us to
protect the system from effective heating through energy conservation. This
avenue, which was not available in the context of superconductors,
effectively provides a means to ``sharpen'' the Fermi step (or even create a
discontinuity at a different momentum), thereby enhancing fermion pairing.

\section{Model}

In our problem, we assume optically trapped bosons in thermal contact with
fermions that occupy two hyperfine states $\left\vert \uparrow \right\rangle
$ and $\left\vert \downarrow \right\rangle $. This system has a Hamiltonian
of the form $\mathcal{\hat{H}}=\mathcal{\hat{H}}_{0}+\mathcal{\hat{H}}_{I}$
where the noninteracting part $\mathcal{\hat{H}}_{0}$ is given by%
\begin{equation*}
\mathcal{\hat{H}}_{0}=\int d^{3}\mathbf{r}\sum_{p}\hat{\psi}_{p}^{\dag }\left( \mathbf{r}\right) \left[ -%
\frac{1}{2m_{p}}\partial ^{2}-\mu _{\alpha }+V_{\alpha }\left( \mathbf{r}%
\right) \right] \hat{\psi}_{p}\left( \mathbf{r}\right) ,
\end{equation*}%
where for brevity, we introduced the subscript $p=B,F_{\uparrow } ,F_{\downarrow } $ which labels bosons ($p=B$) and fermions in the two hyperfine states:
\textquotedblleft up\textquotedblright\ ($p=F_{\uparrow }$) or
\textquotedblleft down\textquotedblright\ ($p=F_{\downarrow }$) and $m_{%
\mathrm{F_{\uparrow }}}=m_{\mathrm{F_{\downarrow }}}\equiv m_{\mathrm{F}}$
is the fermion mass and $m_{B}$ is the mass of the bosons. \ We assume that
fermions in either hyperfine state feel the same trapping potential. \ Thus,
$V_{p}\left( \mathbf{r}\right) $ is given by $V_{F,B}\left( \mathbf{r}%
\right) =\frac{1}{2}m_{F,B}\Omega _{F,B}^{2}r^{2}$ where the subscript $F$ ($%
B$) refers to fermions (bosons). There is also an interaction Hamiltonian $%
\mathcal{{\hat{H}}}_{I}$ which has the form

\begin{equation*}
\mathcal{\hat{H}}_{I}=\frac{1}{2}\int d^{3}\mathbf{r}%
\sum_{p_{1},p_{2}}g_{p_{1},p_{2}}~\hat{\psi}_{p_{1}}^{\dag }\left( \mathbf{r}%
\right) \hat{\psi}_{p_{1}}\left( \mathbf{r}\right) \hat{\psi}_{p_{2}}^{\dag
}\left( \mathbf{r}\right) \hat{\psi}_{p_{2}}\left( \mathbf{r}\right) ,
\end{equation*}

with $g_{p_{1}p_{2}}$ being the strength for $s$-wave collisions between the
particles labelled by $p_{1},p_{2}=\left\{ B,F_{\uparrow },F_{\downarrow
}\right\} $. While Pauli exclusion requires that $g_{F_{\uparrow
}F_{\uparrow }}=g_{F_{\downarrow }F_{\downarrow }}=0$, an attractive
coupling $g_{F_{\uparrow }F_{\downarrow }}\equiv g_{F_{\downarrow
}F_{\uparrow }}<0$ will lead to BCS pairing. A nonzero interaction $g_{FB}$
between bosons and fermions is required for thermalization between the two
populations. We need put no other restrictions on $g_{p_{1}p_{2}},$ but our
desired effect will be easier to observe experimentally with some other
constraints. For instance, requiring that $g_{FB}<0$ will  raise the
BCS condensation temperature \cite{Heiselberg00} while a larger $g_{BB}>0$
facilitates thermalization between bosons and fermions.

To proceed further we use the gap equation, Eq.~(\ref{GapEq}), where we have
$n_{\mathbf{k}}=n_{\mathbf{k}}^{(\mathrm{FD)}}$ at equilibrium. In the
Thomas Fermi approximation, the transition temperature, $T_{\mathrm{c}%
}^{\left( 0\right) }$, is given by \cite{Stoof99}%
\begin{equation}
k_{B}T_{\mathrm{c}}^{\left( 0\right) }\simeq \frac{8\varepsilon
_{F}e^{\gamma -2}}{\pi }\exp \left[ -\frac{\pi }{2k_{F}\left\vert
a_{\uparrow \downarrow }\right\vert }\right] ,  \label{Tc0}
\end{equation}%
where $k_{F}$ is the Fermi wave-vector and $\gamma \approx 0.577\ldots $ is
Euler's constant, the scattering length $a_{\uparrow \downarrow }$ is a
simple combination of the coupling strengths $g_{p_{1}p_{2}}$.

\subsection{Nonequilibrium Enhancement}

It is possible to create distributions that lead to larger order parameter and
effective condensation temperature by weakly perturbing the trapped fermions.
Specifically, we affect a Bragg pulse (Fig.~\ref{Brgpic}) by illuminating
the fermions with two lasers, which are both largely detuned from any
fermionic transition. In what follows, we assume the lasers to be even
further detuned from any bosonic transition such that we may ignore the
effect of the Bragg pulse on the bosons. The interaction of the fermions
with these lasers is described by the addition of a term%
\begin{equation*}
\mathcal{\hat{H}}_{\mathrm{bg}}=\int d^{3}\mathbf{r}\sum_{p_{f}=F_{%
\uparrow },F_{\downarrow }}\hat{\psi}_{p_{f}}^{\dag }\left( \mathbf{r}%
\right) \left[ \hbar\Omega_{bg}\cos \left( \mathbf{q}\cdot \mathbf{r}-\omega
t\right) \right] \hat{\psi}_{p_{f}}\left( \mathbf{r}\right)
\end{equation*}%
to the Hamiltonian where $\mathbf{q}$ and $\omega $ represent the difference
in wavevectors and frequencies between the two lasers \cite{Blakie02}. Now,
following Schmid~\cite{Schmid77} and the general argument in the
introduction [see,~Eq.~(\ref{renorm})], we introduce the function, $\delta
n_{\mathbf{k}}$, which describes departure from equilibrium.  While Eliashberg assumed in \cite{Larkin86} that the impurity concentration was high enough in metals such that momentum relaxation happened at a much faster rate than energy relaxation, we shall not make this assumption.  As such, $\delta
n_{\mathbf{k}}$ need not be isotropic although this requirement is easily included in our model.  The corresponding term, $\chi/\nu(\varepsilon _{\mathrm{F}})$ from Eq.~(\ref{renorm}), is added to the right side of the gap equation~(\ref{GapEq}) and leads
to a new solution, $\Delta >\Delta _{0}$, for the order parameter at the
same temperature (note that in the non-equilibrium situation, the notion of
a temperature of the Fermi system is undefined, and here by temperature we
imply the original temperature and that of the Bose bath). For $%
T-T_{c}^{\left( 0\right) }\ll T_{c}^{\left( 0\right) }$, Eq.~(\ref{GapEq})
can be cast in the form of a Ginzburg-Landau equation, which, including the
nonequilibrium term, becomes%
\begin{equation}
\left( \ln {\frac{T}{T_{c}^{\left( 0\right) }}}-\chi \right) \Delta +b\left(
{\frac{|\Delta |}{{T_{c}^{\left( 0\right) }}}}\right) ^{2}\Delta =0,
\label{GiL}
\end{equation}%
where we assume that the coefficient in the cubic term of Eq.~(\ref{GiL}) is
only weakly affected by the perturbation and use its standard BCS value [$%
\zeta (z)$ is the Riemann zeta-function] $b=7\zeta (3)/(8\pi ^{2})\approx
0.107\ldots $ (see also, Ref.~\cite{Baranov98}). Eq.~(\ref{GiL}) nominally
leads to an exponential enhancement of the effective critical temperature: $%
T_{c}^{\mathrm{eff}}=e^{\chi }T_{c}^{\left( 0\right) }$, if $\chi $ is
positive.

\subsection{Kinetic Equation}

To calculate this enhancement for the Bose-Fermi mixture, we shall balance
the Boltzmann equation for $n_{\mathbf{k}}$ in the spirit of Eliashberg,
including both a collisive contribution and that from Bragg scattering.%
\begin{equation}
\dot{n}_{\mathbf{k}}=I_{\text{coll}}\left[ n_{\mathbf{k}}\right] +I_{\text{%
Bragg}}\left[ n_{\mathbf{k}}\right] .  \label{Boltz}
\end{equation}%
For small departures from equilibrium, we can linearize the collision
integral in $\delta n_{\mathbf{k}}$ and use the $1/\tau $-approximation: $I_{%
\text{coll}}\left[ n\right] =-\delta n/\tau _{0}$, where $\tau _{0}$ is the
quasiparticle lifetime. In our system, this lifetime will be dominated by
the inelastic collision time between bosons and fermions. We may estimate
this time as in \cite{Anderlini05,Capuzzi04} by means of a $%
1/\tau_{0}=n\sigma v$ approximation where $n$ is the boson density at the
center of the trap, $\sigma$ is the constant low temperature cross-section
for boson-fermion scattering, and $v$ is the average relative velocity
associated with the collisions between bosons and fermions.

Note that there exist  other contributions to the collision integral, in particular those coming from fermion collisions.
Our model assumes point-like interactions between fermions: Such interactions can be separated into interactions in the reduced BCS channel, which involve particles with opposite momenta that eventually form Cooper pairs and other types of scattering events, which give rise to Fermi liquid
renormalizations on the high-temperature side and superconducting fluctuations on the BCS side. Note that dropping off the
latter terms would lead to an integrable (Richardson) model that does not have any thermalization processes and therefore
the collision integral for its quasiparticles must vanish. In thermodynamic limit this model is described by BCS mean-field
theory perfectly well and so we can say that the pairing part of fermion interactions is already incorporated in our theory. Of course,
including fermion-boson collision and the second type of fermion interaction processes break integrability and lead to two types
of effects: First, such interactions lead to Fermi liquid renormalizations of the effective mass and the quasiparticle $Z$-factor.
However, these effects are not germane to the physics of interest, and we may assume that all relevant corrections are already included
 and treat our system as that consisting of Fermi liquid quasiparticles. However, there is of a course a second dissipative
part coming from interactions, such as those due to bosons already included into $\tau_0$ and quasiparticle scatterings and decay processes due to
non-BCS fermion-fermion collisions. Similarly to the work of Eliashberg, we will assume that the latter contribution to the collision integral is
less significant than $\tau_0^{-1}$ due to the bosons.  Fermi Liquid quasiparticles are exactly defined precisely on the Fermi sphere, but they have a finite lifetime due to decay processes elsewhere.  By not including the lifetime $\tau_{\mathbf{k}}$ of a quasiparticle at momentum $\mathbf{k}$ in our linearization of the Boltzmann equation, we have implicitly assumed that $\tau_{\mathbf{k}} \gg \tau_{0}$. Because $\frac{1}{\tau_{\mathbf{k}}} \propto (\pi k_{B}T)^{2}+(\varepsilon_{\mathbf{k}}-\varepsilon_{F})^{2}$ in a Fermi Liquid \cite{Pines66}, there will always be an energy region where this assumption will indeed be true for low temperatures.  The most important contribution to the integral in the expression for $\chi$ comes from states for which $\varepsilon_{\mathbf{k}}$ is within $k_{B}T$ of $\varepsilon_{F}$.  As such, if we require that $T \ll T_{F}$ and $\hbar\omega \ll \varepsilon_{F}$, then our linearization of the Boltzmann equation with respect to $\tau_{0}$ will be legitimate for the calculation of an enhancement of superfluidity.  Again, we stress the importance of recognizing that the aforementioned requirements are necessary only for quantitative accuracy of our model.  As with Eliashberg's enhancement of superconductivity, we expect our effect to be observable far outside the constrained parameter space that is necessary for strict validity of our simple model, which provides a proof of principle for using nonequilibrium perturbations to enhance
fermion pairing in cold atom systems.

With these caveats in mind, we shall tune
the frequency of our Bragg pulse such that $\omega\tau_{0} \gg 1$. This will
ensure that any non-stationary part of the distribution function will be
small \cite{Larkin86}. Equivalently, we may think of this requirement as the
statement that the Bragg pulse pumps the system out of equilibrium must
faster than the system relaxes. We may note here that the aforementioned
assumption that $\delta n$ is small also implies that $\chi \ll 1$, thereby
diminishing the desired effect, $T_{c}^{\mathrm{eff}}=(1+\chi )T_{c}^{\left(
0\right)}\sim T_{c}^{\left( 0\right) }$. Again, this approximation greatly
simplifies our theoretical problem by allowing us to expand the Boltzmann equation, but it is only a mathematical convenience that
represents no impediment to an experimentalist looking for striking
enhancements of $T_{c}^{\left( 0\right) }$.

\ Eq.~(\ref{Boltz}) can now be solved for $\delta n$ to yield
\begin{equation}
\delta n_{\mathbf{k}}=\tau _{0}I_{\text{Bragg}}\left[ n_{\mathbf{k}}^{\mathrm{(FD)}}%
\right] \left( 1-e^{-t/\tau _{0}}\right) ,  \label{dn}
\end{equation}%
which shows that a stationary nonequilibrium state is formed in a
characteristic time $\tau _{0}$. The Bragg part in Eq.~(\ref{Boltz}), $I_{%
\text{Bragg}}\left[ n\right] $, now depends only on $n_{\mathbf{k}}^{\mathrm{%
(FD)}}$ and can be computed with Fermi's golden rule. When the wavelength of
the Bragg pulse is much larger than the DeBroglie wavelength of the fermions
and the reciprocal frequency of the pulse is much smaller than the
relaxation time ($\lambda _{F}\left\vert \mathbf{q}\right\vert \ll 1$ and $%
\omega \tau _{0}\gg 1$), Fermi's golden rule yields

\begin{widetext}%
\begin{equation*}
I_{\text{Bragg}}\left[ n_{\mathbf{k}}^{\mathrm{(FD)}}\right] =\frac{2\pi }{%
\hbar }\Omega _{bg}^{2}\left\{ n_{\mathbf{k}-\mathbf{q}}^{\mathrm{(FD)}%
}\left( 1-n_{\mathbf{k}}^{\mathrm{(FD)}}\right) \delta \left( \varepsilon _{%
\mathbf{k}}-\varepsilon _{\mathbf{k}-\mathbf{q}}-\hbar \omega \right) -n_{%
\mathbf{k}}^{\mathrm{(FD)}}\left( 1-n_{\mathbf{k}+\mathbf{q}}^{\mathrm{(FD)}%
}\right) \delta \left( \varepsilon _{\mathbf{k}+\mathbf{q}}-\varepsilon _{%
\mathbf{k}}-\hbar \omega \right) \right\}.
\end{equation*}%
\end{widetext}%

The determination of $I_{\text{Bragg}}\left[ n_{\mathbf{k}}^{F}\right] $
allows us to find $\chi $ and ultimately $T_{c}^{\mathrm{(eff)}}$. The
optimal parameters depend on whether the fermionic gas is in the superfluid
or normal phase at the time the Bragg pulse is applied. In the former case,
the energy gap in the quasiparticle density of states protects a Bragg pulse
with $\hbar \omega <2\Delta _{0}$ from producing new quasiparticles that
will hinder superfluidity. However, in the normal phase, the energy
conservation requirement that $\epsilon _{\mathbf{k}+\mathbf{q}}-\epsilon _{%
\mathbf{k}}=\pm \hbar \omega $ and an independent control of $\mathbf{q}$
and $\omega $ allow us to engineer a pulse that ensures that only
``thermal'' quasiparticles with energies $\varepsilon >\varepsilon_{\mathrm{F%
}}$ are pushed to even higher energies, while the fermions below $%
\varepsilon_{\mathrm{F}}$ are not affected.

Thus far, we have assumed that heat is being dissipated from the fermions
into the bosons through collisions. Because Bose-Einstein condensation
inhibits collisions with fermions and severely reduces thermalization
between the two populations, our simple analysis depends on the bosons being
at a constant temperature $T$ greater than their Bose-Einstein condensation
temperature $T_{BEC}\simeq 0.94\hbar \Omega _{B}N_{B}^{1/3}$ \cite{Albus02}.
Condensation can be prevented at temperatures close to the BCS transition
temperature by having $\Omega _{B}\ll\Omega _{F}$ and $m_{B}\gg m_{F}$.
Treating the bosons classically, we expect their temperature to increase no
faster than $\frac{dT}{dt}=\frac{1}{C}\Omega _{bg}\omega $,
which is the energy pumping rate due to the Bragg pulse for a specific heat $C$.  For a harmonically confined classical gas, we use a specific heat given by $C=3k_{B}N_{B}$.  If $t_{Bragg}$ is
the time over which the Bragg pulse is turned on, then so long as $t_{Bragg}%
\frac{dT}{dt}$ is much less than the temperature of the bosons, we may
consider the bosonic population to be an isothermal bath. This may be
accomplished by having a large number of bosons at low density. One may be
able to avoid the assumption of a bosonic population altogether when driving
the transition from normal to superfluid at temperatures above $%
T_{c}^{\left( 0\right) }$ by allowing energetic particles to leave the trap
as in evaporative cooling. We shall show later that this is possible because
the Bragg pulse may be tuned to couple only to particles with energies above
a threshold energy depending on $\omega$ and $\mathbf{q}$. For concreteness
however, we shall keep the bosonic population throughout the following
section.

\section{Numerical Results}

As an example, we calculate the nonequilibrium enhancement of $T_{c}$ for a
trapped mixture of $^{87}$Rb and $^{6}$Li under the aforementioned
assumptions with $\delta n\ll 1$. We assume a cloud of $10^{5}$ Lithium
atoms and $10^{7}$ Rubidium atoms in traps of frequencies $\Omega
_{F}=200\Omega _{B}=3$ kHz correspondingly. We use scattering lengths $%
a_{BB}=109a_{0}$, $a_{FF}=-2160a_{0}$, $a_{FB}=-100a_{0}$ \cite{Heiselberg00}
and a typical collision time $\tau _{0}\approx 136$ ns as estimated via the $1/\tau_{0}=n\sigma v$ approximation from above. With these parameters, the equilibrium BCS and BEC condensation
temperatures are $T_{c}^{\left( 0\right) }\approx 0.15~T_{F}=291$ nK and $%
T_{BEC}\approx 23.2$ nK.  The quasiparticle lifetime $\tau_{\mathbf{k}}$ for the normal fluid is estimated as $\tau_{\mathbf{k}} \approx 3~\mu$s for $|\varepsilon_{\mathbf{k}}-\varepsilon_{F}| \simeq \hbar\omega$ via the methods in Refs. \cite{Pines66,Shahzamanian02}.

\subsection{Superfluid at Equilibrium}

\begin{figure}[t]
\vspace{-0.1in}
\subfigure[]{\includegraphics[width=2.5in]{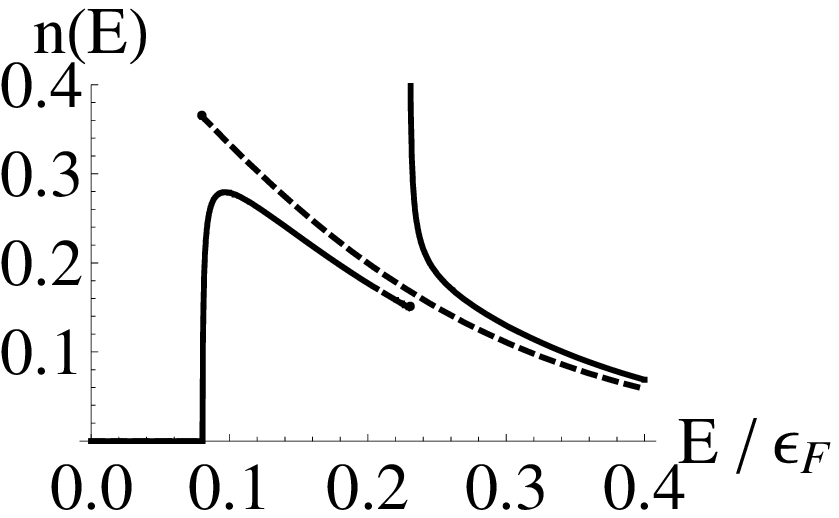}\label{fig:sphaseO1}}
\hspace{0.1in}
\subfigure[]{\includegraphics[width=2.5in]{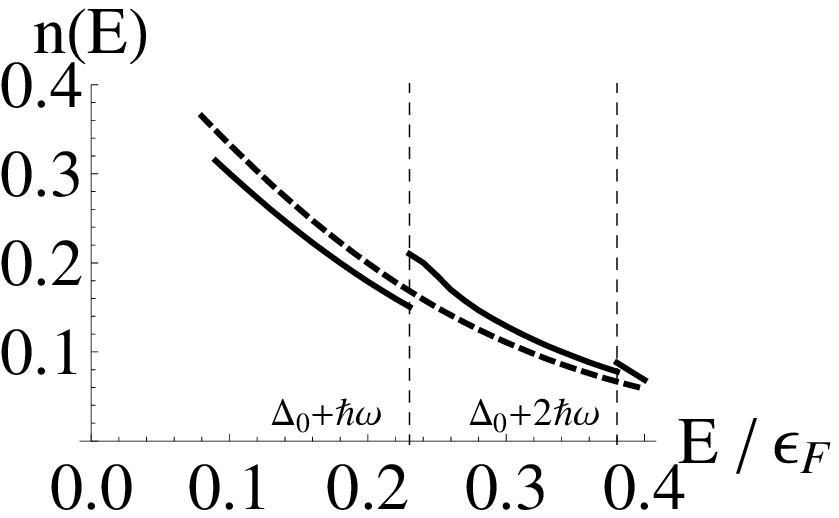}\label{fig:sphase}}
\caption{\protect\footnotesize \subref{fig:sphaseO1} The first order approximation to the quasiparticle occupation as a function of $E=\protect\sqrt{\protect\xi ^{2}+\Delta_0^{2}}$ for parameters $\Omega
_{bg}=70~\Omega _{F}$, $\left\vert \mathbf{q}\right\vert =0.1k_{F}$, and $\hbar \protect\omega =0.15\protect\varepsilon _{F}$.  The unphysical singularities at $E=\Delta_{0}$ and $E=\Delta_{0}+\hbar\omega$ are not included in the calculation of $\chi$.  See text for details.  The equilibrium values for this system are $T=0.14 T_{F}$ and $\Delta_{0}=0.08
\varepsilon_{F}$.  \subref{fig:sphase} The exact distribution function schematically drawn for the same parameters with $\Delta_{0}$ and $\hbar\omega$ in units of $\varepsilon_{F}$.  The thick dashed lines represent the occupation at equilibrium.}
\label{sphase}
\end{figure}

\begin{figure}[t]
\vspace{-0.05in}
\includegraphics[width=2.5in]{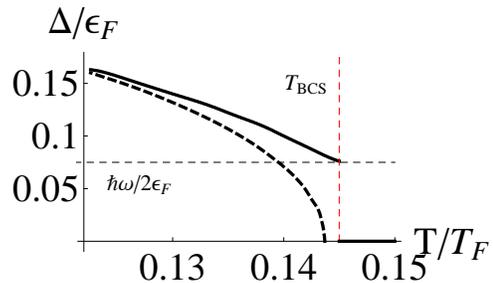}
\caption{\protect\footnotesize The order parameter $\Delta$ as function of the temperature which solves the nonequilibrium gap equation for parameters $\Omega _{bg}=110\Omega _{F}$, $\left\vert \mathbf{q}\right\vert =0.1\ k_{F}$, and $\hbar\omega =0.15\varepsilon_{F}$.  The black dashed line is the equilibrium dependence while the red dashed line gives the nonequilibrium transition temperature $T_{BCS}>T_{BCS}^{\left(0\right)}$.  We have constrained $\Delta_{0} >\hbar\omega/2$ to avoid pair-breaking.}
\label{GapvTsup}
\end{figure}

Let us assume that the system is initially in the superfluid phase at
equilibrium with a low enough temperature $T<T_{c}^{\left( 0\right) }$ such that  $2\Delta > \hbar\omega =0.15\varepsilon_{F}$. Using Eq.~(\ref{dn}), we may calculate the stationary distribution
function (Fig.~\ref{sphase}) for Bragg parameters $\Omega _{bg}=70%
\Omega _{F}$ and $\left\vert \mathbf{q}\right\vert =0.1\ k_{F}$. Some comments on the form of $\delta
n_{\mathbf{k}}$ are necessary. As shown in Fig.~\ref{fig:sphaseO1}, the first order approximation to $n_{\mathbf{k}}$ has unphysical
singularities at $E=\Delta$ and $E=\Delta +\omega $. These are due to first-order transitions
to $E=\Delta +\omega $ from $E=\Delta$ where the quasiparticle density of states
diverges for a superfluid at equilibrium. The exact distribution function, schematically drawn in Fig.~\ref{fig:sphase},
has no infinities.  Higher orders in the expansion of the Boltzmann equation are necessary to curtail the singularities at $\varepsilon=\Delta_{0}+n\omega$ ($n=0,1,2\ldots$).  However, so long as these singularities are localized on energy intervals that are much smaller than $\omega$, the approximate $\delta n_{\mathbf{k}}$ calculated from Eq.~(\ref{dn}) will be suitable for the calculation of both the enhanced order parameter via Eqs. \ref{GapEq} and \ref{renorm} as well as the value of $\chi$ in the Ginzburg-Landau equation (Eq. \ref{GiL}) \cite{Larkin86}.  As expected from the analogy to Eliashberg's work, our singularities have energy widths of about $\Delta \frac{2\pi\tau_{0}\Omega_{bg}^{2}}{\Omega_{F}(6N_{F})^{1/3}}$.  Hence, we may consider the inequality $\frac{\Delta}{\omega}\frac{2\pi\tau_{0}\Omega_{bg}^{2}}{\Omega_{F}(6N_{F})^{1/3}} \ll 1$ as a further requirement for the validity of our linearization of the Boltzmann equation for the calculation of $\chi$ in the superfluid case.  We may also note that due to the fact that $\hbar \omega <2\Delta _{0}$, no new quasiparticles are excited
from the lower branch by pair breaking. Hence, the quasiparticle number is conserved in this first order approximation ($\sum_{\mathbf{k}}\delta n_{\mathbf{k}}=0$).  The quasiparticles are simply redistributed from the gap edge to higher energies. Substituting $\delta n_{\mathbf{k}}$ into Eq.~(\ref{renorm}), we find that at $T\simeq 0.13$ $T_{F}$ we calculate an increase in $\Delta $ by a factor of $1/10$. So long as $T<T_{c}^{\left( 0\right) }$, the
relative enhancement increases with temperature because there are more
particles to redistribute and the pulse does not have enough energy to break
Cooper pairs.  Unlike in the normal fluid where $\Delta_{0}=0$ at equilibrium, the enhancement $\chi$ depends on the initial value of the gap.  In Fig.~\ref{GapvTsup}, we plot the temperature dependence of the enhanced nonequilibrium gap with a slightly stronger pulse given by $\Omega_{bg}=110\Omega_{F}$, $\hbar\omega =0.15\varepsilon_{F}$, and $\left\vert \mathbf{q}\right\vert =0.1\ k_{F}$.  Note that after the pulse is applied at equilibrium, the temperature of the bath may be increased above $T_{c}^{\left( 0\right) }$ while maintaining a nonzero gap.  This temperature dependence is exactly what would be expected from the analogous plot in Ref. \cite{Larkin86}.  The new BCS transition temperature $T_{c}$ is given by the maximum of the nonequilibrium plot where the inequality $\hbar \omega <2\Delta _{0}$ is saturated.  If the temperature is further increased, $\Delta$ discontinuously vanishes again.  For these parameters we calculate a small 3\% increase of the transition temperature as expected from our requirement that nonlinear effects (even gap enhancing effects) be ignored in the Boltzmann equation.  The temperatures we have considered are sufficiently close to the equilibrium transition temperature such that the approximation $T_{BCS}\approx T_{BCS}^{\left(0\right)}(1+\chi)$ is justified.  For small $\Delta_{0}$ a crude, order of magnitude estimate of $\chi$ may be given by $\chi \sim \frac{2\pi\hbar\Omega_{bg}^{2}\tau_{0}}{\varepsilon_{F}}n_{0}(\Delta_{0})n_{0}(\Delta_{0}+\hbar\omega)$.

\subsection{Normal at Equilibrium}

\begin{figure}[t]
\includegraphics[width=2.5in]{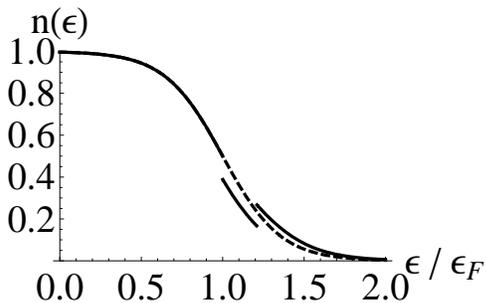}
\caption{{\protect\footnotesize The }$T=0.18~T_{F}${\protect\footnotesize \
normal phase occupation as a function of energy for }$\Omega
_{bg}=110~\Omega _{F}${\protect\footnotesize , }$\left\vert \mathbf{q}%
\right\vert =0.1~k_{F}${\protect\footnotesize , and }$\hbar \protect\omega %
=0.21\protect\varepsilon _{F}${\protect\footnotesize . \ The enhanced
nonequilibrium transition temperature is $T_{BCS}%
=1.3~T_{BCS}^{\left( 0\right) }$ The dashed line is the occupation at
equilibrium.}}
\label{nphase}
\end{figure}

\begin{figure}[!t]
\includegraphics[width=2.5in]{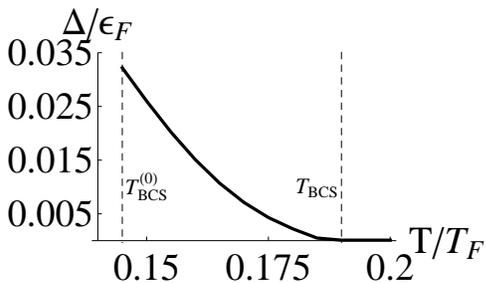}
\caption{\protect\footnotesize The enhanced gap $\Delta$ as function of temperatures above $T_{BCS}^{\left(0\right)}$ for fermions in the normal phase at equilibrium and parameters $\Omega_{bg}=110~\Omega _{F}$, $\left\vert \mathbf{q}\right\vert =0.1k_{F}$, and $\hbar\omega=0.21~\varepsilon_{F}$}
\label{GvTnorm}
\end{figure}

For contrast, let us now assume that we start in the normal phase at equilibrium.  The initial distribution function is simply the Fermi-Dirac function for the noninteracting quasiparticles of Fermi Liquid theory.  We have $\Delta _{0}=0$, so Eliashberg's requirement that $\hbar \omega < 2\Delta _{0}$ cannot be satisfied. However, we can guarantee that particles are not
excited from below the Fermi level by choosing $\omega $ and $\mathbf{q}$
such that the constraint $4\varepsilon _{\mathbf{q}}\varepsilon _{F}\leq
\left( \hbar \omega -\varepsilon _{\mathbf{q}}\right) ^{2}$ is enforced.  Because of this requirement, momentum and energy conservation cannot be simultaneously achieved for particles with energies less than $\varepsilon_{F}$.
As such, only particles outside the Fermi sphere can undergo transitions.  The lower equilibrium occupation number and higher density of states at high energies ensures that quasiparticles just outside the Fermi sphere are excited to higher energies.

Although in our system we have particle conservation just as in the superfluid case, we comment here that if these excited particles are allowed to leave the trap, then we have effectively cooled
the fermions by sharpening the Fermi step, and the boson bath is
unnecessary. Substituting $\delta n_{\mathbf{k}}$ into Eq.~(\ref{renorm}), we see that the depression of the population at the Fermi level
shown in Fig.~\ref{nphase} allows for a nonzero $\Delta $ above $%
T_{BCS}^{\left( 0\right) }$. As the temperature is increased, the nonequilibrium gap enhancement is overpowered by thermal smearing of the distribution function.  There are more quasiparticles at energies near $\varepsilon_{F}$ where they most strongly hinder superfluidity.  This contrasts with the superfluid situation wherein the enhancement increases with temperature so long as $T<T_{BCS}^{\left(0\right)}$.  This effect can be seen in Fig.~\ref{GvTnorm}, where we find an enhancement of $T_{BCS}$ by about 30\% for parameters $\Omega _{bg}=110~\Omega _{F}$, $\left\vert \mathbf{q}\right\vert =0.1\ k_{F}$, and $\hbar \omega=0.21\varepsilon _{F}$.  This increase is much more drastic than the enhancement in the superfluid phase because the requirements that $\delta n_{\mathbf{k}} \ll n_{\mathbf{k}}^{\mathrm{(FD)}}$ and $\hbar\omega \ll \varepsilon_{F}$ are much less stringent than the superfluid requirements $\frac{\Delta}{\omega}\frac{2\pi\tau_{0}\Omega_{bg}^{2}}{\Omega_{F}(6N_{F})^{1/3}} \ll 1$ and $\hbar\omega \le \Delta$.  Thus, we may use a stronger pulse while still linearizing the Boltzmann equation legitimately.  As such, fermionic
superfluidity is expected to appear at temperatures as high as $T\approx
1.3~T_{c}^{\left( 0\right) }$.

\section{Summary}

To conclude, we have shown that perturbing a system of trapped fermions
creating a stationary quasiparticle distribution can be an effective way to
stimulate fermion pairing and superfluidity. To demonstrate this, we
calculate enhancements of the BCS order parameter and the transition
temperature for a mixture of $^{87}$Rb and $^{6}$Li that is pushed out of
equilibrium by a Bragg pulse. The mechanism by which fermions within the
Fermi sphere are not excited differs depending on initial conditions. If the
gas is a superfluid at equilibrium, these excitations are precluded by
keeping $\hbar \omega <2\Delta _{0}$. In the normal phase, the parameters of
the pulse can be chosen such that fermions below a certain energy cannot 
simultaneously satisfy momentum and energy conservation.  Thus, they are not
excited. In both cases, the enhancements that we calculate are small, but
this is a consequence of our perturbative treatment rather than a physical
constraint. This is evidenced by the strong effects reported from
experiments on superconductors~\cite{Blamire91,Heslinga93}, which are based
on the same underlying mechanism. Finally, we suggest that by enhancing or
creating a discontinuity in the quasistationary strongly non-equilibrium
distribution of fermions (not necessarily at the Fermi momentum) via the
technique proposed in this paper, one may achieve effective BCS pairing at
nominally very high temperatures of the bath.

The authors are grateful to Roman Barankov, Leo Radzihovsky, Gil Refael,
Jeff Rogers, and Ian Spielman for illuminating discussions of this topic.
A.R. acknowledges the National Defense Science and Engineering Graduate
(NDSEG) Fellowship. This research is supported by DARPA and US-ARO.

\bibliographystyle{apsrev}
\bibliography{NeqEnhance}

\begin{thebibliography}{28}
\expandafter\ifx\csname natexlab\endcsname\relax\def\natexlab#1{#1}\fi
\expandafter\ifx\csname bibnamefont\endcsname\relax
  \def\bibnamefont#1{#1}\fi
\expandafter\ifx\csname bibfnamefont\endcsname\relax
  \def\bibfnamefont#1{#1}\fi
\expandafter\ifx\csname citenamefont\endcsname\relax
  \def\citenamefont#1{#1}\fi
\expandafter\ifx\csname url\endcsname\relax
  \def\url#1{\texttt{#1}}\fi
\expandafter\ifx\csname urlprefix\endcsname\relax\def\urlprefix{URL }\fi
\providecommand{\bibinfo}[2]{#2}
\providecommand{\eprint}[2][]{\url{#2}}

\bibitem[{\citenamefont{Regal et~al.}()\citenamefont{Regal, Greiner, and
  Jin}}]{CABCS}
\bibinfo{author}{\bibfnamefont{C.~A.} \bibnamefont{Regal}},
  \bibinfo{author}{\bibfnamefont{M.}~\bibnamefont{Greiner}}, \bibnamefont{and}
  \bibinfo{author}{\bibfnamefont{D.~S.} \bibnamefont{Jin}},
  \bibinfo{howpublished}{Phys. Rev. Lett. {\bf 92}, 040403 (2004); M. Greiner
  and C. A. Regal and J. T. Stewart and D. S. Jin, {\em ibid.} {\bf 94}, 110401
  (2005); C.H. Schunck, M.W. Zwierlein, A. Schirotzek, and W. Ketterle, {\em
  ibid.} {\bf 98}, 050404 (2007)}.

\bibitem[{\citenamefont{Holland et~al.}(2001)\citenamefont{Holland, Kokkelmans,
  Chiofalo, and Walser}}]{Holland01}
\bibinfo{author}{\bibfnamefont{M.}~\bibnamefont{Holland}},
  \bibinfo{author}{\bibfnamefont{S.~J. J. M.~F.} \bibnamefont{Kokkelmans}},
  \bibinfo{author}{\bibfnamefont{M.~L.} \bibnamefont{Chiofalo}},
  \bibnamefont{and} \bibinfo{author}{\bibfnamefont{R.}~\bibnamefont{Walser}},
  \bibinfo{journal}{Phys. Rev. Lett.} \textbf{\bibinfo{volume}{87}},
  \bibinfo{pages}{120406} (\bibinfo{year}{2001}).

\bibitem[{\citenamefont{Stoof et~al.}(1996)\citenamefont{Stoof, Houbiers,
  Sackett, and Hulet}}]{Stoof96}
\bibinfo{author}{\bibfnamefont{H.~T.~C.} \bibnamefont{Stoof}},
  \bibinfo{author}{\bibfnamefont{M.}~\bibnamefont{Houbiers}},
  \bibinfo{author}{\bibfnamefont{C.~A.} \bibnamefont{Sackett}},
  \bibnamefont{and} \bibinfo{author}{\bibfnamefont{R.~G.} \bibnamefont{Hulet}},
  \bibinfo{journal}{Phys. Rev. Lett.} \textbf{\bibinfo{volume}{76}},
  \bibinfo{pages}{10} (\bibinfo{year}{1996}).

\bibitem[{\citenamefont{Zwierlein et~al.}()\citenamefont{Zwierlein, Schunck,
  Schirotzek, and Ketterle}}]{Zwierlein06}
\bibinfo{author}{\bibfnamefont{M.~W.} \bibnamefont{Zwierlein}},
  \bibinfo{author}{\bibfnamefont{C.~H.} \bibnamefont{Schunck}},
  \bibinfo{author}{\bibfnamefont{A.}~\bibnamefont{Schirotzek}},
  \bibnamefont{and} \bibinfo{author}{\bibfnamefont{W.}~\bibnamefont{Ketterle}},
  \bibinfo{howpublished}{Nature {\bf 442}, 54-58 (2006)}.

\bibitem[{\citenamefont{Inguscio et~al.}()}]{Stoof99}
\bibinfo{author}{\bibfnamefont{M.}~\bibnamefont{Inguscio}}
  \bibnamefont{et~al.}, \bibinfo{howpublished}{Bose-Einstein Condensation in
  Atomic Gases, Enrico Fermi Summer School, Course CXL, IOS Press, Amsterdam
  (1999)}.

\bibitem[{\citenamefont{Wyatt et~al.}(1966)\citenamefont{Wyatt, Dmitriev,
  Moore, and Sheard}}]{Wyatt66}
\bibinfo{author}{\bibfnamefont{A.~F.~G.} \bibnamefont{Wyatt}},
  \bibinfo{author}{\bibfnamefont{V.~M.} \bibnamefont{Dmitriev}},
  \bibinfo{author}{\bibfnamefont{W.~S.} \bibnamefont{Moore}}, \bibnamefont{and}
  \bibinfo{author}{\bibfnamefont{F.~W.} \bibnamefont{Sheard}},
  \bibinfo{journal}{Phys. Rev. Lett.} \textbf{\bibinfo{volume}{16}},
  \bibinfo{pages}{1166} (\bibinfo{year}{1966}).

\bibitem[{\citenamefont{Eliashberg}({\natexlab{a}})}]{Eliashberg70}
\bibinfo{author}{\bibfnamefont{G.~M.} \bibnamefont{Eliashberg}},
  \bibinfo{howpublished}{Pis'ma Zh. Eksp. Teor. Fiz. {\bf 11}, 186 (1970); JETP
  Lett. {\bf 11}, 114 (1970)}.

\bibitem[{\citenamefont{Eliashberg}({\natexlab{b}})}]{Larkin86}
\bibinfo{author}{\bibfnamefont{G.~M.} \bibnamefont{Eliashberg}},
  \emph{\bibinfo{title}{\textit{Nonequilibrium Superconductivity}}},
  \bibinfo{howpublished}{edited by D. N. Langenberg and A. I. Larkin
  (North-Holland, New York, 1986)}.

\bibitem[{\citenamefont{Tredwell and Jacobsen}(1976)}]{Tredwell75}
\bibinfo{author}{\bibfnamefont{T.~J.} \bibnamefont{Tredwell}} \bibnamefont{and}
  \bibinfo{author}{\bibfnamefont{E.~H.} \bibnamefont{Jacobsen}},
  \bibinfo{journal}{Phys. Rev. B} \textbf{\bibinfo{volume}{13}},
  \bibinfo{pages}{2931} (\bibinfo{year}{1976}).

\bibitem[{\citenamefont{Pals and Dobben}(1979)}]{Pals79}
\bibinfo{author}{\bibfnamefont{J.~A.} \bibnamefont{Pals}} \bibnamefont{and}
  \bibinfo{author}{\bibfnamefont{J.}~\bibnamefont{Dobben}},
  \bibinfo{journal}{Phys. Rev. B} \textbf{\bibinfo{volume}{20}},
  \bibinfo{pages}{935} (\bibinfo{year}{1979}).

\bibitem[{\citenamefont{Chang and Scalapino}()}]{Chang78}
\bibinfo{author}{\bibfnamefont{J.}~\bibnamefont{Chang}} \bibnamefont{and}
  \bibinfo{author}{\bibfnamefont{D.~J.} \bibnamefont{Scalapino}},
  \bibinfo{howpublished}{J. Low Temp. Phys. {\bf31}, 1-32 (1978)}.

\bibitem[{\citenamefont{Blamire et~al.}(1991)\citenamefont{Blamire, Kirk,
  Evetts, and Klapwijk}}]{Blamire91}
\bibinfo{author}{\bibfnamefont{M.~G.} \bibnamefont{Blamire}},
  \bibinfo{author}{\bibfnamefont{E.~C.~G.} \bibnamefont{Kirk}},
  \bibinfo{author}{\bibfnamefont{J.~E.} \bibnamefont{Evetts}},
  \bibnamefont{and} \bibinfo{author}{\bibfnamefont{T.~M.}
  \bibnamefont{Klapwijk}}, \bibinfo{journal}{Phys. Rev. Lett.}
  \textbf{\bibinfo{volume}{66}}, \bibinfo{pages}{220} (\bibinfo{year}{1991}).

\bibitem[{\citenamefont{Heslinga and Klapwijk}(1993)}]{Heslinga93}
\bibinfo{author}{\bibfnamefont{D.~R.} \bibnamefont{Heslinga}} \bibnamefont{and}
  \bibinfo{author}{\bibfnamefont{T.~M.} \bibnamefont{Klapwijk}},
  \bibinfo{journal}{Phys. Rev. B} \textbf{\bibinfo{volume}{47}},
  \bibinfo{pages}{5157} (\bibinfo{year}{1993}).

\bibitem[{\citenamefont{Carr et~al.}(2004)\citenamefont{Carr, Shlyapnikov, and
  Castin}}]{Carr04}
\bibinfo{author}{\bibfnamefont{L.~D.} \bibnamefont{Carr}},
  \bibinfo{author}{\bibfnamefont{G.~V.} \bibnamefont{Shlyapnikov}},
  \bibnamefont{and} \bibinfo{author}{\bibfnamefont{Y.}~\bibnamefont{Castin}},
  \bibinfo{journal}{Phys. Rev. Lett.} \textbf{\bibinfo{volume}{92}},
  \bibinfo{pages}{150404} (\bibinfo{year}{2004}).

\bibitem[{\citenamefont{Bruun and Clark}()}]{Bruun00}
\bibinfo{author}{\bibfnamefont{G.~M.} \bibnamefont{Bruun}} \bibnamefont{and}
  \bibinfo{author}{\bibfnamefont{C.~W.} \bibnamefont{Clark}},
  \bibinfo{howpublished}{J. Phys. B. {\bf33}, 3953-3959 (1978)}.

\bibitem[{\citenamefont{S\'a~de Melo et~al.}(1993)\citenamefont{S\'a~de Melo,
  Randeria, and Engelbrecht}}]{Sademelo93}
\bibinfo{author}{\bibfnamefont{C.~A.~R.} \bibnamefont{S\'a~de Melo}},
  \bibinfo{author}{\bibfnamefont{M.}~\bibnamefont{Randeria}}, \bibnamefont{and}
  \bibinfo{author}{\bibfnamefont{J.~R.} \bibnamefont{Engelbrecht}},
  \bibinfo{journal}{Phys. Rev. Lett.} \textbf{\bibinfo{volume}{71}},
  \bibinfo{pages}{3202} (\bibinfo{year}{1993}).

\bibitem[{\citenamefont{Sheehy and Radzihovsky}(2007)}]{Sheehy06}
\bibinfo{author}{\bibfnamefont{D.~E.} \bibnamefont{Sheehy}} \bibnamefont{and}
  \bibinfo{author}{\bibfnamefont{L.}~\bibnamefont{Radzihovsky}},
  \bibinfo{journal}{Annals of Physics} \textbf{\bibinfo{volume}{322}},
  \bibinfo{pages}{1790 } (\bibinfo{year}{2007}).

\bibitem[{\citenamefont{Blakie et~al.}(2002)\citenamefont{Blakie, Ballagh, and
  Gardiner}}]{Blakie02}
\bibinfo{author}{\bibfnamefont{P.~B.} \bibnamefont{Blakie}},
  \bibinfo{author}{\bibfnamefont{R.~J.} \bibnamefont{Ballagh}},
  \bibnamefont{and} \bibinfo{author}{\bibfnamefont{C.~W.}
  \bibnamefont{Gardiner}}, \bibinfo{journal}{Phys. Rev. A}
  \textbf{\bibinfo{volume}{65}}, \bibinfo{pages}{033602}
  (\bibinfo{year}{2002}).

\bibitem[{\citenamefont{Rey et~al.}(2005)\citenamefont{Rey, Blakie, Pupillo,
  Williams, and Clark}}]{Rey05}
\bibinfo{author}{\bibfnamefont{A.~M.} \bibnamefont{Rey}},
  \bibinfo{author}{\bibfnamefont{P.~B.} \bibnamefont{Blakie}},
  \bibinfo{author}{\bibfnamefont{G.}~\bibnamefont{Pupillo}},
  \bibinfo{author}{\bibfnamefont{C.~J.} \bibnamefont{Williams}},
  \bibnamefont{and} \bibinfo{author}{\bibfnamefont{C.~W.} \bibnamefont{Clark}},
  \bibinfo{journal}{Phys. Rev. A} \textbf{\bibinfo{volume}{72}},
  \bibinfo{pages}{023407} (\bibinfo{year}{2005}).

\bibitem[{\citenamefont{Butts and Rokhsar}(1997)}]{Butts97}
\bibinfo{author}{\bibfnamefont{D.~A.} \bibnamefont{Butts}} \bibnamefont{and}
  \bibinfo{author}{\bibfnamefont{D.~S.} \bibnamefont{Rokhsar}},
  \bibinfo{journal}{Phys. Rev. A} \textbf{\bibinfo{volume}{55}},
  \bibinfo{pages}{4346} (\bibinfo{year}{1997}).

\bibitem[{\citenamefont{Heiselberg et~al.}(2000)\citenamefont{Heiselberg,
  Pethick, Smith, and Viverit}}]{Heiselberg00}
\bibinfo{author}{\bibfnamefont{H.}~\bibnamefont{Heiselberg}},
  \bibinfo{author}{\bibfnamefont{C.~J.} \bibnamefont{Pethick}},
  \bibinfo{author}{\bibfnamefont{H.}~\bibnamefont{Smith}}, \bibnamefont{and}
  \bibinfo{author}{\bibfnamefont{L.}~\bibnamefont{Viverit}},
  \bibinfo{journal}{Phys. Rev. Lett.} \textbf{\bibinfo{volume}{85}},
  \bibinfo{pages}{2418} (\bibinfo{year}{2000}).

\bibitem[{\citenamefont{Schmid}(1977)}]{Schmid77}
\bibinfo{author}{\bibfnamefont{A.}~\bibnamefont{Schmid}},
  \bibinfo{journal}{Phys. Rev. Lett.} \textbf{\bibinfo{volume}{38}},
  \bibinfo{pages}{922} (\bibinfo{year}{1977}).

\bibitem[{\citenamefont{Baranov and Petrov}(1998)}]{Baranov98}
\bibinfo{author}{\bibfnamefont{M.~A.} \bibnamefont{Baranov}} \bibnamefont{and}
  \bibinfo{author}{\bibfnamefont{D.~S.} \bibnamefont{Petrov}},
  \bibinfo{journal}{Phys. Rev. A} \textbf{\bibinfo{volume}{58}},
  \bibinfo{pages}{R801} (\bibinfo{year}{1998}).

\bibitem[{\citenamefont{Anderlini et~al.}(2005)\citenamefont{Anderlini,
  Ciampini, Cossart, Courtade, Cristiani, Sias, Morsch, and
  Arimondo}}]{Anderlini05}
\bibinfo{author}{\bibfnamefont{M.}~\bibnamefont{Anderlini}},
  \bibinfo{author}{\bibfnamefont{D.}~\bibnamefont{Ciampini}},
  \bibinfo{author}{\bibfnamefont{D.}~\bibnamefont{Cossart}},
  \bibinfo{author}{\bibfnamefont{E.}~\bibnamefont{Courtade}},
  \bibinfo{author}{\bibfnamefont{M.}~\bibnamefont{Cristiani}},
  \bibinfo{author}{\bibfnamefont{C.}~\bibnamefont{Sias}},
  \bibinfo{author}{\bibfnamefont{O.}~\bibnamefont{Morsch}}, \bibnamefont{and}
  \bibinfo{author}{\bibfnamefont{E.}~\bibnamefont{Arimondo}},
  \bibinfo{journal}{Phys. Rev. A} \textbf{\bibinfo{volume}{72}},
  \bibinfo{pages}{033408} (\bibinfo{year}{2005}).

\bibitem[{\citenamefont{Capuzzi et~al.}(2004)\citenamefont{Capuzzi, Vignolo,
  Toschi, Succi, and Tosi}}]{Capuzzi04}
\bibinfo{author}{\bibfnamefont{P.}~\bibnamefont{Capuzzi}},
  \bibinfo{author}{\bibfnamefont{P.}~\bibnamefont{Vignolo}},
  \bibinfo{author}{\bibfnamefont{F.}~\bibnamefont{Toschi}},
  \bibinfo{author}{\bibfnamefont{S.}~\bibnamefont{Succi}}, \bibnamefont{and}
  \bibinfo{author}{\bibfnamefont{M.~P.} \bibnamefont{Tosi}},
  \bibinfo{journal}{Phys. Rev. A} \textbf{\bibinfo{volume}{70}},
  \bibinfo{pages}{043623} (\bibinfo{year}{2004}).

\bibitem[{\citenamefont{Pines and Nozieres}(1966)}]{Pines66}
\bibinfo{author}{\bibfnamefont{D.}~\bibnamefont{Pines}} \bibnamefont{and}
  \bibinfo{author}{\bibfnamefont{P.}~\bibnamefont{Nozieres}},
  \emph{\bibinfo{title}{Quantum Liquids}} (\bibinfo{publisher}{W.A. Benjamin},
  \bibinfo{address}{New York}, \bibinfo{year}{1966}).

\bibitem[{\citenamefont{Albus et~al.}()\citenamefont{Albus, Giorgini,
  Illuminati, and Viverit}}]{Albus02}
\bibinfo{author}{\bibfnamefont{A.~P.} \bibnamefont{Albus}},
  \bibinfo{author}{\bibfnamefont{S.}~\bibnamefont{Giorgini}},
  \bibinfo{author}{\bibfnamefont{F.}~\bibnamefont{Illuminati}},
  \bibnamefont{and} \bibinfo{author}{\bibfnamefont{L.}~\bibnamefont{Viverit}},
  \bibinfo{howpublished}{J. Phys. B. {\bf35}, L511-L519 (2002)}.

\bibitem[{\citenamefont{Shahzamanian and Yavary}(2002)}]{Shahzamanian02}
\bibinfo{author}{\bibfnamefont{M.~A.} \bibnamefont{Shahzamanian}}
  \bibnamefont{and} \bibinfo{author}{\bibfnamefont{H.}~\bibnamefont{Yavary}},
  \bibinfo{journal}{Physica B: Condensed Matter}
  \textbf{\bibinfo{volume}{321}}, \bibinfo{pages}{385 } (\bibinfo{year}{2002}),
  ISSN \bibinfo{issn}{0921-4526}.

\end{thebibliography}

\end{document}